\documentclass[12pt]{article}
\usepackage{epsfig}
\usepackage{amsfonts}
\begin {document}

\title {NEWTONIAN GRAVITATIONAL MULTIPOLES AS GROUP-INVARIANT SOLUTIONS}

\author{J.L. Hern\'andez-Pastora\thanks{E.T.S. Ingenier\'\i a
Industrial de B\'ejar. Phone: +34 923 408080 Ext
2263. Also at +34 923 294400 Ext 1308. e-mail address: jlhp@usal.es}\\
\\
Departamento de Matem\'atica Aplicada. \\ Universidad de Salamanca.  Salamanca,
Espa\~na.  }

\date{} \maketitle

\begin{abstract}
A family of vector fields that  are the infinitesimal generators of determined one-parameter groups of transformations are constructed. It is shown that these vector fields represent symmetries of the system of differential equations interrelated by  the axially symmetric Laplace equation  and a certain supplementary equation. Group-invariant solutions of this system of equations are obtained by means of two alternative methods, and it is proved that these solutions turn out  to be the family of axisymmetric potentials related to specific gravitational multipoles. The existence of these symmetries provides us with a generalization of the fact that the Newtonian Monopole is defined by the solution of the Laplace equation  with spherical symmetry, and it allows us to extract from all solutions of this equation those with the prescribed Newtonian Multipole Moments.

\end{abstract}


\newpage

\section{Introduction}

In Newtonian gravity the gravitational field of a  bounded mass distribution is given as a solution of the  Poisson equation. As is known, that solution can be expanded in a power series of the
inverse  radial coordinate in the neighbourhood of  infinity, in such a way that some suitable quantities, referred to as Multipole Moments (MM), characterize the specific solution. These (MM) quantities are related to the structure of the mass distribution by means of integrals defined over the source, and hence the solution provides a well defined and physically meaningful description of Newtonian gravity.

When addressing the description of gravity in  the vacuum, we look for solutions of the Laplace equation, and the above-mentioned multipole expansion allows us to identify the arbitrary constants of the general solution in terms of the Multipole Moments (MM) of the source. Thus, the general well-behavioured solution of the Laplace equation at infinity (or far enough from the source) is a series expansion in powers of $1/r$ the terms of which represent the different contributions of each MM that provide both  the physical characterisctics of the source and a description of the exterior gravitational field.

Let us  stress two relevant features of this description of the gravitational solutions: first, because of the linearity of the Laplace equation that series can be seen as the sum  of exact solutions, and hence the partial sums of that series become a new solution of the equation. We refer to these partial sums as the Multipole Solutions of Newtonian gravity.
 
Secondly, we wish to emphasize that the definition of  MM gives them a well known physical meaning  in terms of certain integrals defined over the mass distribution.

These characteristics allow us to consider the multipole  expansion series as a perturbative approach to  the interpretation of the gravitational field  by following two criteria: the description of the field is performed with respect to a point sufficiently far away from the source that higher powers of the inverse radial coordinate can be neglected; alternatively, we truncate the multipole expansion series at a suitable Multipole Solution because the mass distribution of the source provides infinitesimal higher-order contributions.

Therefore,  MM play a  fundamental role in the construction of solutions of  Newtonian gravity since  Multipole Solutions can be understood as corrections in the desciption of the gravitational field of  the first term of the multipole expansion (the Monopole solution). This first term of the succession of partial sums represents the gravitational field generated by a bounded spherical source.
The spherical  feature characterizes  that specific solution univocally and, in fact, if the  spherical symmetry condition is imposed on the solutions of the Laplace equation, then  a unique solution of the series is obtained:  the gravitational monopole. One way to introduce that constraint is by requiring the solution to be invariant under the action of the symmetry group $SO(3)$. In spherical coordinates $\lbrace r, \theta,\varphi \rbrace$, the infinitesimal generators of that group are:
\begin{eqnarray}
&\textbf{J}_3&=\frac{\partial}{\partial \varphi}\\ \nonumber
&\textbf{J}_2&=-\cos \varphi \frac{\partial}{\partial \theta}+ \sin \varphi \cot \theta \frac{\partial}{\partial \varphi}\\ \nonumber
&\textbf{J}_1&=\sin\varphi \frac{\partial}{\partial \theta}+ \cos \varphi \cot \theta \frac{\partial}{\partial \varphi} \ .
\end{eqnarray}

The invariance condition over a gravitational potential  $\phi$ leads to the equations $\displaystyle{\frac{\partial\phi}{\partial \varphi}=0}$, $\displaystyle{\frac{\partial\phi}{\partial \theta}=0}$, and therefore  the Laplace equation reduces to the following equation
\begin{equation}
\frac{\partial}{\partial r}\left( r^2 \frac{\partial \phi}{\partial r}\right) =0 \ ,
\end{equation}
or, equivalently, 
\begin{equation}
2 \phi_r+r\phi_{rr}=0 \ ,
\label{symeq}
\end{equation}
whose unique solution (different from the trivial constant solution) is simply the gravitational monopole $\phi=\displaystyle{-\frac{a}{r}}$, (with the imposed condition $a > 0$, from the demand of positive mass). 

This is the point of departure in the line of enquiry pursued here: is there any kind of symmetry that describes  gravitational multipole solutions? Might it be possible to find some differential equations such as (\ref{symeq}) that restrict the solutions of the Laplace equation to those with the prescribed multipole characteristic? 
The spherical symmetry, in the way described above, becomes a universal symmetry in the sense that it does not depend on the equation we wish to solve namely the Laplace equation. This is not our aim for the symmetries we are seeking, but the claim is the existence of some group of transformations that will be a symmetry of the Laplace equation,  which allows us to choose a specific kind of solution from the general equation. 

In what follows we shall see that it is possible to construct a family of vector fields that are the infinitesimal generators of a one-parameter group of transformations. Furthermore, this group turns out to be a symmetry of a system of equations consisting of the Laplace equation and  another equation, henceforth  called {\it supplementary} equation.

Thus, by means of the symmetry, the resolution of the Laplace equation when we are looking for multipole solutions simplifies the system of equations to be solved, and the problem proves to be easier,  since we only  need to integrate the supplementary equation (which is a very simple  ODE in the angular variable alone), choosing those given by the correspondent invariance condition for the integration functions.

This work is organized as follows: in section $2$, along a first subsection  we offer a brief review of  gravitational multipole solutions in Newtonian gravity, and we recall what multipole moments are and their physical meaning. In order to avoid too many complicated formulae in the main text, we have used an appendix to explain details of these contents.

In a second subsection we find  some differential equations that are fulfilled by the partial sums of the series that describe  gravitational multipole solutions. These equations represent a family of additional conditions to the axially symmetric Laplace equation, acting as a restriction to its general solution.

In section $3$ we construct a family of vector fields that play the role of infinitesimal generators of some one-parameter group of transformations. It is proved that these groups are symmetries of a system of equations consisting of  the Laplace equation and a specific supplementary equation. In a first step, we show the so-called Monopole-Dipole symmetry and then we generalize the symmetry to the $2^N$-pole order moment case,  where the notion of generalized symmetries must be introduced.

Section $4$ is devoted to calculating the groups of transformations for these  vector fields that  generate the symmetries. Finally, in section $5$ we show that the calculation of group-invariant solutions of the axially symmetric Laplace equation leads to the gravitational multipole solutions of Newtonian gravity.

A Conclusion section contains some comments about  the aims achieved and possible future  generalizations.

Finally, two appendices are included: Appendix A contains information about Classical Multipole Moments, and Appendix B is devoted to complete the calculation of the functions involved in the construction of the symmetries.
\section{Multipole solutions in Newtonian gravity}

\subsection{The multipole moments and gravitational multipole solution}

In Newtonian gravity we need to introduce completely symmetric and trace-free tensor fields to define the   so-called Classical Multipole Moments\footnote{The definition and introduction of Classical Multipole Moments is explained with details in the Appendix A}. As can be seen in the appendix A, for the case of axial symmetry only one component of each tensor is independent and the exterior gravitational field generated by an axially symmetric source of mass density $\rho(z,\theta)$  is given by the following potential
\begin{equation}
\phi=-G \sum_{n=0}^\infty\frac{M_n}{r^{n+1}}P_n(\cos\theta) \ ,
\label{serie}
\end{equation}
where $(r,\theta)$ are the  radial and polar coordinates of the exterior point with respect to any origin on the axes of symmetry; $P_n$ are  Legendre polynomials and the constants
 $M_n$ are the multipole moments (\ref{dieciseis}) of the source.

Just as the  Poisson equation applies for the description of gravity generated by a mass distribution, so does the Laplace equation  for the vacuum case. Thus, if we consider both cases in our configuration space   $M$, the problem to solve could be divided into different domains ($M=\Omega_{\rho}\cup\Omega_v$) for each region as follows:
\begin{equation}
\bigtriangleup\phi=
\left\{
\begin{array}{c}
\rho(\vec z) \qquad ,   \qquad \vec x \in \Omega_{\rho}\\
0 \qquad ,   \qquad \vec x \in \Omega_{v} \qquad ,
\end{array}
\right.
\end{equation}
$\Omega_{\rho}$, $\Omega_v$ being the domains of the region containing all the mass distributions  and vacuum regions respectively.
In the $\Omega_v$ domain, the general solution with axial symmetry  and good asymptotic  behaviour (i.e. decaying to zero at infinity or far from the source)
\begin{equation}
\bigtriangleup\phi= \frac{1}{r^2}\frac{\partial}{\partial r}\left( r^2 \frac{\partial\phi}{\partial r}\right) +\frac{1}{r^2\sin\theta}\frac{\partial}{\partial \theta}\left(\sin\theta \frac{\partial\phi}{\partial \theta}\right)=0 
\label{laplace}
\end{equation}
is given by the series:
\begin{equation}
\phi=\sum_{n=0}^\infty\frac{a_n}{r^{n+1}}P_n(\cos\theta) \ ,
\label{vacuum}
\end{equation}
where $a_n$ are arbitrary constants without any physical meaning. Nevertheless, since we must demand  continuity of the function $\phi$ at the boundary of regions $\Omega_{\rho}$ and $\Omega_v$, then these constants $a_n$ are (apart from the constant $G$) identical to the multipole moments $M_n$ and acquire the meaning of those quantities.

\subsection{Some properties of  multipole solutions}

We shall now show some properties of the series (\ref{serie}). The derivative of $\phi$ with respect to  the variable $r$ proves to be (let us take $G=1$):
\begin{eqnarray}
\phi_r=\sum_{n=0}^{\infty} (n+1) \frac{M_n}{r^{n+2}} P_n(y) = - \frac{\phi}{r}+\frac{1}{r}\sum_{n=0}^{\infty} n \lambda_n P_n(y) \ , \ \displaystyle{\lambda_n\equiv \frac{M_n}{r^{n+1}}}\\ \nonumber
y\equiv\cos\theta \ ,
\end{eqnarray}
and hence we have the following relation:
\begin{eqnarray}
r \phi_r+\phi=R_1 \label{rphi}\\ 
R_1 \equiv \sum_{n=1}^{\infty} n \lambda_n P_n(y) \label{rphiR1}
 \ .
\end{eqnarray}
The above expression can be read as a linear first-order differential equation for the potential $\phi$; it is clear that since $R_1$, the right part of the equation (\ref{rphi}), does not depend on  $\lambda_0$, the first term of the series (\ref{serie}) is a solution of the corresponding homogeneous differential equation.

The derivative of $\phi$ with respect to the variable $y$ is:
\begin{equation}
\phi_y=-\sum_{n=0}^{\infty} \lambda_n \partial_y P_n(y)= -\lambda_1-\sum_{n=2}^{\infty} \lambda_n \partial_y P_n(y) \ ,
\end{equation}
and hence the following relation holds:
\begin{eqnarray}
r \phi_r+\phi+y \phi_y=\sum_{n=2}^{\infty} n \lambda_n P_n(y)-\sum_{n=2}^{\infty} y \lambda_n \partial_y P_n(y) = R_2 \label{ryphi}\\
R_2 \equiv - \sum_{n=2}^{\infty}  \lambda_n \partial_y P_{n-1}(y) \ , \label{ryphiR2}
\end{eqnarray}
by means of the property of the Legendre polynomials $nP_n(y)+P_{n-1}^{\prime}(y)=yP_n^{\prime}(y)$.
By again  considering   expression (\ref{ryphi}) as a linear first-order differential equation for the potential $\phi$, we  realize that since $R_2$ (\ref{ryphiR2}) does not depend on $\lambda_1$,  the partial sum of order 2 (i.e., the sum of the first two terms) of the series (\ref{serie})  is a solution of the corresponding homogeneous differential equation (\ref{ryphi}). 

Moreover, since the first term of $R_2$ (\ref{ryphiR2}) is $-\lambda_2$, we can add to the first member of the equation (\ref{ryphi}) the second derivative (multiplied by a suitable factor) of $\phi$ with respect to the variable $y$  (see equation (\ref{deriy})) in order to obtain a new differential  equation for $\phi$ with an independent term of higher order on $\lambda_n$. By means of a recursive calculation, and considering successive derivatives of $\phi$ with respect to the variable $y$,
\begin{equation}
\frac{\partial^i \phi}{\partial y^i} = -\sum_{n=i}^{\infty} \lambda_n \frac{\partial^iP_n(y)}{\partial y^i}=-\lambda_i i! L_{i,i} -\sum_{n=i+1}^{\infty} \lambda_n \sum_{k=i}^{n} \frac{k!}{(k-i)!} y^{k-i}L_{n,k} \ ,
\label{deriy}
\end{equation}
where $L_{n,k}$ are the coefficients of the Legendre polynomial of degree $n$ for the power $k$ of its variable $y$, we can obtain the following differential equation for any value of $N \geq 2$:
\begin{equation}
r\phi_r+\phi+y \phi_y+\sum_{n=2}^N h_n(y) \partial^n_y \phi=R_{N+1} \ ,
\label{rn}
\end{equation}
where  $R_{N+1}$ is a series on $\lambda_n$ of  order\footnote{The order of the series must be understood as the smallest value of the index $n$ for a non-null term $\lambda_n C_{k,n}$ of the series. Thus, the sum (\ref{erek}) starts at the value of that order. (Note that $C_{1,0}(y)=0$ and $C_{2,1}(y)=0$ since $C_{1,n}(y)=nP_n(y)$ and $C_{2,n}(y)=-\partial_y P_{n-1}(y)$).} $\lambda_{N+1}$, and $h_n(y)$ are unknown polynomials  of $y$ that we need to introduce to obtain an independent term $R_{N+1}$ of that order.  In fact, starting from (\ref{rphiR1}) and (\ref{ryphiR2}), $R_k$ (with $k \geq 1$) are series on $\lambda_n$ which can be written as follows:
\begin{equation}
R_k=\sum_{n=k}^{\infty} \lambda_n C_{k,n}(y) \ ,
\label{erek}
\end{equation}
that is to say, it can be seen  that, for all $k\geq 1$, $C_{k,n}(y)=0$ 
with $n<k$; in particular, by using (\ref{deriy}) and (\ref{erek}) (note that $L_{i,i}=(2i-1)!!/i!$) one can quickly conclude that if one choses $R_k$ recursively in the following way:
\begin{equation}
R_{k+1}=R_{k}+\frac{C_{k,k}(y)}{(2k-1)!!} \partial_y^k \phi \ , 
\label{erres}
\end{equation}
then, the $\lambda_k$ term of each $R_{k+1}$ will vanish for all $k\geq 1$, and therefore from equation (\ref{rn}) the functions $h_n(y)$  are given by:
\begin{equation}
h_n(y)=\frac{C_{n,n}(y)}{n! L_{n,n}}=\frac{C_{n,n}(y)}{(2n-1)!!} \ .
\label{hach}
\end{equation}
In appendix B we develop the recurrence relation (\ref{erres}) in order to obtain the degree and coefficients of the polynomials $h_n(y)$, which is needed for the construction of the symmetries and the proof of theorems in  following sections.

\vskip 2mm

In conclusion, we have that a solution of the corresponding homogeneous equation (\ref{rn}) for any $N$ is the partial sum of order $N+1$ of the series (\ref{serie}). Therefore, the  homogeneous part of equation (\ref{rn}) can be considered as a supplementary condition that must satisfy  a solution of the Laplace equation in order to be the gravitational multipole solution of  order $N$, or, in  other words, can be understood as a condition to truncate the series (\ref{serie}).

\section{Multipole symmetries}

Let 
\begin{equation}
\textbf{v}= \xi(x,u)\frac{\partial}{\partial r}+\tau(x,u) \frac{\partial}{\partial y}+\sigma(x,u)\frac{\partial}{\partial u}
\label{uve}
\end{equation}
be a vector field  on an open subset $M \subset X \times U$, where $X={\mathbb {R}}^2$, with coordinates $x=(r,y)$, is the space representing the independent variables, and $U={\mathbb {R}}$,  with coordinate $u$, that  represents the dependent variable.

Let  $\bigtriangleup_{\nu}(x,u^{(n)})=0$ be a system of  $\nu$ differential equations defined over $M$. As  is known \cite{olvers}, if $G$ is  a local group  of transformations acting on $M$ and 
\begin{equation}
pr^{(n)}\textbf{v}\left[ \bigtriangleup_{\nu}(x,u^{(n)})\right]  =0 \ ,
\label{pro}
\end{equation}
whenever $\bigtriangleup_{\nu}(x,u^{(n)})=0$, for every infinitesimal generator $\textbf{v}$ of $G$, then $G$ is a symmetry group of the system, where the vector field $pr^{(n)}\textbf{v}$ is the $n$-th prolongation of $\textbf{v}$ defined on the corresponding jet space $M^{(n)}\subset X\times U^{(n)}$, (whose coordinates represent the independent variables, the dependent variable, and the derivatives  of the dependent variable up to order $n$)
\begin{equation}
pr^{(n)}\textbf{v}=\textbf{v}+\sum_J \sigma^J(x,u^{(n)})\frac{\partial}{\partial u_J} \ ,
\label{provn}
\end{equation}
where the summation being over all multi-indices $J=(j_1,...,j_k)$, with $1\leq j_k\leq 2$, $1\leq k\leq n$, and the notation $u_J \equiv \displaystyle{\frac{\partial u}{\partial x^J}}$ is used.

A scalar function $\Psi(x,u)=0$ on $M$ is an invariant of the vector field $\textbf{v}$ if the following equation is fulfilled
\begin{equation}
\xi(x,u)\frac{\partial \Psi}{\partial r}+\tau(x,u) \frac{\partial \Psi}{\partial y}+\sigma(x,u)\frac{\partial \Psi}{\partial u}=0 \ ,
\label{inv}
\end{equation}
 and taking into account that $D_r(\Psi)=0=\displaystyle{\frac{\partial \Psi}{\partial u}\frac{\partial
 u}{\partial r}+\frac{\partial \Psi}{\partial r}}$ and
 $D_y(\Psi)=0=\displaystyle{\frac{\partial\Psi}{\partial u}\frac{\partial
 u}{\partial y}+\frac{\partial \Psi}{\partial y}}$, (where $D$ denotes total derivative) then  equation 
 (\ref{inv}) can be written in terms of the function $u(x)$ as follows:
\begin{equation}
\xi(x,u)\frac{\partial u}{\partial r}+\tau(x,u) \frac{\partial u}{\partial y}-\sigma(x,u)=0 \ .
\label{invu}
\end{equation}

\subsection{The Monopole-Dipole symmetry}

Let \begin{equation}
\textbf{v}= r\frac{\partial}{\partial r}+y \frac{\partial}{\partial y}-u\frac{\partial}{\partial u}
\label{uvemd}
\end{equation}
be a vector field  considered as the infinitesimal generator  of some group $G_{MD}$. Looking for invariants $\Psi(r,y,u)$ of this vector field (i.e: $r\Psi_r+y\Psi_y-u\Psi_u=0$) we end up with the following condition for the variable $u$:
\begin{equation}
r u_r+y u_y+u = 0
\end{equation}
 which is merely the homogeneous equation (\ref{ryphi}) satisfied by  the gravitational potential  describing
  the monopole-dipole solution of the Laplace equation. 

We shall now check  whether the one-parameter group generated by the infinitesimal generator $\bf v$ (\ref{uvemd}) is  a symmetry group of the Laplace equation. In order to do so, we first calculate the second prolongation (\ref{provn}) of $\bf v$ by means of  formula (see formula (2.39) in \cite{olvers} for details)
\begin{equation}
\sigma^J(x,u^{(2)})=D_J\left(\sigma(x,u)-r u_r-y u_y\right)+r u_{J,r}+y u_{J,y} \ ,
\label{Dtotal}
\end{equation}
where $u_{J,x}\equiv \displaystyle{\frac{\partial u_J}{\partial x}}$, 
and we have the following:
\begin{equation}
pr^{(2)} \textbf{v} = \textbf{v}- \displaystyle{2 u_r \frac{\partial}{\partial u_{r}}-2 u_y \frac{\partial}{\partial u_{y}}-3 u_{rr}\frac{\partial}{\partial u_{rr}}-3 u_{yy} \frac{\partial}{\partial u_{yy}}-3 u_{ry} \frac{\partial}{\partial u_{ry}}} \ .
\end{equation}
If this is applied to the Laplace equation (with axial symmetry)  
\begin{equation}
\bigtriangleup u \equiv r^2u_{rr}+2 r u_r-2 y u_y+(1-y^2) u_{yy}=0 \ ,
\label{laplacey}
\end{equation}
 we obtain:
\begin{equation}
pr^{(2)} \textbf{v} \left[\bigtriangleup u \right]= -\left[\bigtriangleup u\right]-2 u_{yy} \ .
\label{2uyy}
\end{equation}
Therefore, we can state the following 

\noindent {\bf Theorem 1}:

The system of equations $\bigtriangleup_{\nu}(x,u^{(n)})=0$ given by
\begin{equation}
\left\{
\begin{array}{c}
\bigtriangleup_1(x,u^{(n)})  \equiv  \bigtriangleup u=0 \\ 
\bigtriangleup_2(x,u^{(n)})  \equiv  u_{yy}=0
\end{array}
\right.
\label{system}
\end{equation}
where $\bigtriangleup_1$ being the Laplace equation (\ref{laplacey}) and $\bigtriangleup_2$ the so called supplementary equation, admits a symmetry group whose infinitesimal generator is {\bf v}.
\vspace*{5mm}

\noindent Proof:

Since  the second prolongation of {\bf v} acting on the supplementary equation is $pr^{(2)} \textbf{v}\left[\bigtriangleup_2\right]=-3 u_{yy}$, and according to (\ref{2uyy}) we have $pr^{(2)}\textbf{v} \left[\bigtriangleup_1\right]=0=pr^ {(2)}\textbf{v}\left[\bigtriangleup_2\right]$ whenever $\bigtriangleup_{\nu}(x,u^{(n)})=0$, we conclude with the proof. $\hfill{\square}$

\vspace*{5mm}

\subsection{$2^N$-pole-order symmetry}

The homogeneous differential equation from expression 	(\ref{rn}) is the equation that must be fulfilled by the gravitational multipole solution up to $2^N$-pole-order. In analogy to the previous case, we now introduce a new vector field and  claim that the condition satisfied by an invariant function of that vector field on $M$ reproduces the homogeneous equation (\ref{rn}). This kind of vector field is as follows:
\begin{equation}
\textbf{v}=r\displaystyle{\frac{\partial}{\partial r}+y\frac{\partial}{\partial y}-\left[u+\sum_{n=2}^{N}h_n(y) \partial_y^n u\right]\frac{\partial}{\partial u}}
\label{uveN}
\end{equation}
As can be seen immediately  in the above expression, the coefficient function of the $\displaystyle{\frac{\partial}{\partial u}}$ derivative at $\textbf{v}$ depends not only on $x$ and $u$ but also  on the derivatives of $u$. This leads to a significant generalization of the notion of symmetry group, obtained by relaxing the geometric assumption that as long as the coefficient function of the vector field depends on $x$ and $u$ such dependence will generate a (local) one-parameter group of transformation acting on the underlying space $M$. Noether was the first to recognize that it is possible to significantly extend the application of symmetry group methods by including derivatives of the dependent variables in the infinitesimal generators of the transformations\footnote{A complete discussion of the curious history of generalized symmetries can be found in \cite{olvers}}.

Henceforth, we use the term {\it generalized} vector field  to refer to that  kind of vector field  (\ref{uveN}) and we shall denote by $\sigma\left[ u \right]=\sigma\left[ x,u^{(n)} \right]$ any smooth differential function depending on $x$, $u$ and derivatives of $u$ up to order $n$ defined for $(x,u^{(n)}) \in M^{(n)}\subset X \times U^{(n)}$.

Given a generalized vector field such as (\ref{uveN}), its infinite prolongation is the formally infinite sum:
\begin{equation}
pr \textbf{v}=r\displaystyle{\frac{\partial}{\partial r}+y\frac{\partial}{\partial y}+\sum_J \sigma^J\frac{\partial}{\partial u_J}} \ ,
\label{prouveN}
\end{equation}
where $\sigma^J$ is given by (\ref{Dtotal}). As  is known, a generalized vector field $\textbf{v}$ is a generalized infinitesimal symmetry \cite{olvers} of a system of $\nu$ differential equations $\bigtriangleup_{\nu}(x,u^{(n)})=0$ if and only if $pr \textbf{v}\left[\bigtriangleup_{\nu}\right]=0$ for every smooth solution $u=f(x)$, in  direct analogy with the infinitesimal symmetry criterion given previously.

The generalized vector field $\textbf{v}$ (\ref{uveN}) has  the prolongation
\begin{equation}
pr \textbf{v}=\textbf{v}+\sigma^r \displaystyle{\frac{\partial}{\partial u_r}+\sigma^y \frac{\partial}{\partial u_y}+\sigma^{rr} \frac{\partial}{\partial u_{rr}}+\sigma^{yy} \frac{\partial}{\partial u_{yy}}+ \cdots} \ ,
\end{equation}
where the only relevant components for our case are (note that in the following expressions the prime sign is used to  denote derivation of functions $h_n(y)$ with respect to variable $y$)
\begin{eqnarray}
\sigma^r&=&-2 u_r-\sum_{n=2}^{N} h_n(y)  \partial_y^n u_r\\ 
\sigma^y&=&-2 u_y-\sum_{n=2}^N \left(h_n(y)\partial_y^{n+1} u +h_n^{\prime}(y)\partial_y^n u\right)\\
\sigma^{rr}&=&-3 u_{rr}-\sum_{n=2}^Nh_n(y)\partial_y^n u_{rr}\\
\sigma^{yy}&=&-3 u_{yy}-\sum_{n=2}^N \left(h_n(y)\partial_y^{n+2} u+2 h_n^{\prime}(y) \partial_y^{n+1} u+h_n^{\prime \prime}(y) \partial_y^n u\right)
\end{eqnarray}
By applying this prolongation to the Laplace equation (\ref{laplacey}), we obtain:
\begin{eqnarray}
pr \textbf{v}\left[\bigtriangleup_1\right]&=&-\bigtriangleup_1[u]-2 u_{yy} +\\ \nonumber
& & -\sum_{n=2}^N h_n(y)\left(2 r \partial_y^n u_r-2 y \partial_y^{n+1}u+r^2 \partial_y^n u_{rr}+ (1-y^2) \partial_y^{n+2}u \right)+\\ \nonumber
& &-\sum_{n=2}^N h_n^{\prime}(y)\left(-2 y \partial_y^n u+(1-y^2) 2 \partial_y^{n+1}u \right)+h_n^{\prime \prime}(y) (1-y^2) \partial_y^n u
\end{eqnarray}
 It is straightforward to see that:
\begin{eqnarray}
pr \textbf{v}\left[\bigtriangleup_1\right]&=&-\bigtriangleup_1\left[u\right]-2 u_{yy} +\\ \nonumber
& &-\sum_{n=2}^N h_n(y)\left(D_y^n\left[\bigtriangleup_1 \left[u\right]\right]+n(n+1)\partial_y^n u+2 n y \partial_y^{n+1} u\right)+\\ \nonumber
& & -\sum_{n=2}^Nh_n^{\prime}(y)\left(-2 y \partial_y^n u+(1-y^2) 2 \partial_y^{n+1}u \right)+h_n^{\prime \prime}(y) (1-y^2) \partial_y^n u
\end{eqnarray}
and whenever $\bigtriangleup_1 \left[ u \right]=0$ (i.e., over all smooth solutions $u=f(x)$ of the Laplace equation), we have:
\begin{equation}
pr \textbf{v}\left[\bigtriangleup_1 \left[u\right] \right]=-2 u_{yy} -\displaystyle{\sum_{n=2}^N \left[ \Omega_n(y) \partial_y^n u +\Pi_n(y) \partial_y^{n+1} u \right]} \ ,
\label{omypi}
\end{equation}
where
\begin{eqnarray}
\Omega_n(y)&\equiv&n(n+1) h_n(y)-2 y h_n^{\prime}(y)+(1-y^2) h_n^{\prime \prime}(y)\\
\Pi_n(y)&\equiv&2 n y h_n(y)+2 (1-y^2)h_n^{\prime}(y) \ .
\label{accesoria}
\end{eqnarray}
From the above expressions (\ref{omypi}), and considering expressions (\ref{haches}), (\ref{hachexp}), for the functions $h_n(y)$, it can be seen  that:
\begin{equation}
\Omega_k(y)+\Pi_{k-1}(y) = 0 \qquad , \qquad   3\leq k \leq N \ .
\label{omypirel}
\end{equation}
That is to say, it can be proved that the functions $h_n(y)$ defined by (\ref{haches}) satisfy (for $k \geq 3$): 
\begin{equation}
k(k+1) h_k(y) -2 y h_k^{\prime}(y)+ (1-y^2) h_k^{\prime \prime}+2 (k-1) y h_{k-1}(y)+2 (1-y^2) h_{k-1}^{\prime}(y)=0
\end{equation}
And thus, since $\Omega_2(y)=-2$, for every vector field $\textbf{v}$ (\ref{uveN}) we have calculated from (\ref{omypi})-(\ref{omypirel}) that:
\begin{equation}
pr \textbf{v}\left[\bigtriangleup_1 \left[u\right] \right]= -\Pi_N \partial_y^{N+1} u \ , \qquad N \geq 2 \ ,
\end{equation}
and therefore, since $\Pi_N \equiv 2 N y h_N(y)+2 (1-y^2) h_N^{\prime}(y) \neq 0$, we have that  $pr \textbf{v}\left[\bigtriangleup_1\right]=0$ if and only if
\begin{equation}
\displaystyle{\frac{\partial^{N+1} u}{\partial y^{N+1}}=0} \ ,
\label{accesN}
\end{equation}
an equation that we shall call {\it supplementary} equation and denote by $\bigtriangleup_2\left[u\right]$.

With all these results we can state the following:

\noindent {\bf Theorem 2}:

The generalized vector field $\textbf{v}$ (\ref{uveN}) is a generalized infinitesimal symmetry of the system of equations $\bigtriangleup_{\nu}(x,u^{(n)})=0$ given by:
\begin{equation}
\left\{
\begin{array}{c}
\bigtriangleup_1(x,u^{(n)}) \equiv \bigtriangleup u=0 \\ 
\bigtriangleup_2^{(N)}(x,u^{(n)}) \equiv \partial_y^{N+1} u=0 
\end{array}
\right.
\label{systemN}
\end{equation}
\vspace*{5mm}

\noindent Proof:

We shall show that  for every smooth solution $u=f(x)$ of the system we have $pr\textbf{v}\left[\bigtriangleup_{\nu}\right]=0$.

From equation (\ref{accesN}), it is obvious that $pr \textbf{v}\left[\bigtriangleup_1 \left[u\right]\right]=0$. Now, with respect to the supplementary equation we have that 
\begin{equation}
pr \textbf{v} \left[\bigtriangleup_2^{(N)} \left[u\right]\right] = \displaystyle{\sigma^{\overbrace{y \cdots y}^{N+1}}} \ ,
\end{equation}
where $\displaystyle{\sigma^{\overbrace{y \cdots y}^{N+1}}}$ is the coefficient of the prolongation of $\textbf{v}$ defined as follows
\begin{equation}
\displaystyle{\sigma^{\overbrace{y \cdots y}^{N+1}}}=D_y^{N+1}\left(-u-\sum_{n=2}^N h_n(y)\partial_y^n u - r u_r-y u_y\right)+r \partial_y^{N+1} u_r+y \partial_y^{N+2} u
\end{equation}
Accordingly, by developing the total derivative we have:
\begin{equation}
\displaystyle{\sigma^{\overbrace{y \cdots y}^{N+1}}}=-(N+2) \partial_y^{N+1} u -\sum_{k=0}^{N+1} \displaystyle{{{N+1} \choose {k}}\sum_{n=2}^{N}  \partial_y^{N+1-k} h_n(y) \partial_y^{n+k} u} \ .
\label{eso}
\end{equation}
The first term at the right-hand side of (\ref{eso}) is proportional to the left part of equation $\bigtriangleup_2^{(N)}\left[u\right]$ and therefore, it will be zero for all solution $u$ of the system (\ref{systemN}). Moreover, we can see that all  derivatives of $u$ with respect to $y$ appearing in the sums of (\ref{eso}) are higher than $N+3$, because, as we already know,  the functions $h_n(y)$ are polynomials of degree $n-2$ with respect to $y$ and hence all  derivatives $\displaystyle{\partial_y^{N+1-k} h_n(y)}$ vanish for $N+1-k > n-2$, or equivalently $n+k < N+3$.

For these reasons, the derivatives of $u$ with respect to $y$ appearing in (\ref{eso}) can be written as total derivatives of $\bigtriangleup_2^{(N)} \left[u\right]$, and therefore, we can finally write the prolongation acting on $\bigtriangleup_2^{(N)}\left[u\right]$ as follows:
\begin{eqnarray}
pr \textbf{v}\left[\bigtriangleup_2^{(N)}\left[u\right]\right]&=&-(N+2) \left[\bigtriangleup_2^{(N)} \left[u\right]\right] + \label{prolo}\\
& &-\sum_{k=0}^{N+1} \displaystyle{{{N+1} \choose {k}}\sum_{n=2}^{N}  \partial_y^{N+1-k} h_n(y) D_y^{n+k-N-1}\left[\bigtriangleup_2^{(N)}\left[u\right]\right]}.\nonumber
\end{eqnarray}

With this argument, it is clear that (\ref{prolo}) is zero whenever $\bigtriangleup_2^{(N)} u=0$, and so we conclude the proof. $\hfill{\square}$

\vspace*{5mm}

\section{The group of transformations}

Given a vector field $\textbf{v}$, the parametrized maximal integral curve $\Upsilon(\epsilon, x)$ passing through $x$ in $M$ is called the flow generated by $\textbf{v}$. As  is known, the flow is exactly  the same as a local group action on the manifold $M$ or the so-called one-parameter group of transformation.

Therefore, the vector field $\textbf{v}$ is called the infinitesimal generator of the action, since by Taylor's theorem in local coordinates we have that:
\begin{equation}
\Upsilon(\epsilon, x) \equiv exp(\epsilon \textbf{v})(x) = x+\epsilon \xi(x)+O(\epsilon^2)  ,
\end{equation}
where $\xi=(\xi^1, \cdots, \xi^m)$ are the coefficients of $\textbf{v}$, and the notation referred to as exponentiation of the vector field is used. The orbits of the one-parameter group action are the maximal integral curves of the vector field and there is a one-to-one correspondence between local one-parameter groups of transformations and their infinitesimal generators.

\subsection{Monopole-Dipole group $G_{MD}$}

The computation of the  flow  generated by the vector field $\textbf{v}$ (\ref{uve}), which is the infinitesimal generator of the Monopole-Dipole symmetry, is done by solving the following system of ordinary differential equations: 
\begin{eqnarray}
r &=&\displaystyle{\frac{d}{d \epsilon} (e^{\epsilon} r)\mid_{\epsilon=0} } \\ \nonumber
y &=&\displaystyle{\frac{d}{d \epsilon} (e^{\epsilon} y)\mid_{\epsilon=0} } \\ \nonumber
-u &=&\displaystyle{\frac{d}{d \epsilon} (e^{-\epsilon} u)\mid_{\epsilon=0}  } \quad ,
\end{eqnarray}
which leads to 
\begin{equation}
exp(\epsilon \textbf{v})(x,u) =\left(e^{\epsilon} r, e^{\epsilon} y, e^{-\epsilon} u\right) \ ,
\end{equation}

Thus, the one-parameter group $G_{MD}$ generated by $\textbf{v}$ of (\ref{uve}) is given by the following transformation:
\begin{equation}
\left(\tilde{x}, \tilde{u} \right)=\left(e^{\epsilon} r, e^{\epsilon} y, e^{-\epsilon} u\right) \ ;
\label{action}
\end{equation}
i.e., this symmetry group is a kind of scaling symmetry on $M\subseteq X \times U$.

Since the group $G_{MD}$ was proved to be a symmetry group, if $u=f(x)$ is a solution of the system of differential equations (\ref{system}), so are the functions:
\begin{equation}
\tilde{u}=e^{-\epsilon} u = e^{-\epsilon} f(r,y) = e^{-\epsilon} f(e^{-\epsilon} \tilde{r}, e^{-\epsilon} \tilde{y}) \ .
\label{trans}
\end{equation}

\subsection{$2^N$-pole-order groups $G_{M_n}^{(N)}$}

With respect to the generalized vector field $\textbf{v}$ (\ref{uveN}),  some comments should be made about its related group of transformations. First, its one-parameter group can no longer act geometrically on the underlying domain $M$ because the coefficients of $\textbf{v}$ depend on derivatives of $u$, which are also being transformed. Nor can we define a prolonged group action on any finite jet space $M^{(n)}$, since the coefficients of $pr^{(n)} \textbf{v}$ will depend on still higher order derivatives of $u$ than appear in $M^{(n)}$. As  is known, the best way to resolve this is to define an action of the group on a space  of smooth functions as follows \cite{olvers}:
\begin{equation}
\left[ exp(\epsilon \textbf{v}_Q) f\right](x) \equiv u(x,\epsilon) \ ,
\label{actionvq}
\end{equation}
where $u(x,\epsilon)$ is the solution (provided it exists) to the Cauchy problem  of the system of evolution equations:
\begin{eqnarray}
\displaystyle{ \frac{\partial u}{\partial \epsilon}= Q(x, u^{(m)})} &\nonumber \\ 
u(x,0)=f(x)&\ ,
\label{cochi}
\end{eqnarray}
$Q(x, u^{(m)})\equiv Q\left[u\right]$ being the so-called {\it characteristic} of  the {\it evolutionary vector field} $\textbf{v}_Q$ defined as follows \cite{olvers}, \cite{gaeta}:
\begin{equation}
\textbf{v}_Q \equiv Q\left[u\right] \displaystyle{\frac{\partial}{\partial u}} \ ,
\end{equation}
and associated with any generalized vector field (see formula (5.7) in \cite{olvers} for details). In our case, the characteristic of the evolutionary vector field associated with (\ref{uveN}) is 
\begin{equation}
Q\left[u\right] = -u -\sum_{n=2}^N h_n(y) \partial_y^n u -r u_r - y u_y \ .
\end{equation}

As is known \cite{olvers}, if $P\left[u\right]$ is any differential function, and $u(x,\epsilon)$ a smooth solution to (\ref{cochi}), then  the prolongation of the evolutionary vector field determines the infinitesimal change in $P$ under the one-parameter group generated by $\textbf{v}_Q$\footnote{i.e., from the definition of the group action (\ref{actionvq}) we have that $P\displaystyle{\left[exp(\epsilon \textbf{v}_Q) f \right]=P\left[ f \right]+\epsilon pr\textbf{v}_Q (P)\left[f\right]+O(\epsilon^2)}$, since $\displaystyle{\frac{d}{d\epsilon} P\left[u\right]=pr \textbf{v}_Q(P)}$.}. If we assume convergence of the entire Taylor series in $\epsilon$ of the group action, we obtain the  Lie series:
\begin{equation}
P\displaystyle{\left[exp(\epsilon \textbf{v}_Q) f \right]=\sum_{n=0}^{\infty} \frac{\epsilon^n}{n!}  (pr\textbf{v}_Q)^n P\left[ f \right]} \ ,
\label{desa}
\end{equation}
where the power $n$ at the prolongation denotes the sucessive application of $pr \textbf{v}_Q$  $n$ times. In particular, if $P\left[u\right]=u$ then (\ref{desa}) provides the formal series solution to the evolutionary system (\ref{cochi}):
\begin{equation}
u(x, \epsilon)= f(x)+\epsilon pr \textbf{v}_Q(f)+\frac 12 \epsilon^2 (pr \textbf{v}_Q)^2 (f)+O(\epsilon^3) \ ,
\label{infini}
\end{equation}
where, by means of the formula (see (5.6) in \cite{olvers}) $pr \textbf{v}_Q=\displaystyle{\sum_J D_J Q\left[u \right] \frac{\partial}{\partial u_J}}$, we have that
\begin{eqnarray}
 pr \textbf{v}_Q (f)&=& Q\left[u\right]\mid_f \label{esto} \\
 (pr \textbf{v}_Q)^2 (f) &\equiv & pr \textbf{v}_Q (Q)\mid_f =  \left(-Q-\sum_{n=2}^N h_n(y) D_y^n Q-r D_r 
Q- y D_y Q\right)\mid_f \ .\nonumber
\end{eqnarray}

This formal series (\ref{infini}) is no longer of practical use, since we are forced to assume that the solution to the Cauchy problem (\ref{cochi}) is uniquely determined provided the initial data  $f(x)$ is chosen in some appropriate space of functions. Accordingly, the resulting flow will be on the above function space. Verification of this hypothesis involves overcoming a very difficult problem regarding the existence and uniqueness of solutions. And moreover, (\ref{infini}) does not give  an explicit expression of the complete transformation group but the infinitesimal one, and, as was above mentioned, one has to assume its convergence.

Nevertheless, we have presented here these results because, as we shall see in the following section,  Lie series solution  (\ref{infini}), (\ref{esto}), provides us with an argument to introduce a procedure for calculating the $G_{M_n}^{(N)}$ group-invariant solutions by considering those solutions of the system of differential equations that make the characteristic $Q$  zero.

\vspace*{5mm}

An alternative method for calculating the transformation group is now approached, as follows \cite{ibragimov}.
Let us consider the infinite prolongation  (\ref{prouveN}) of the generalized vector field $\textbf{v}$ and let us write down the infinite system of ordinary differential equations:
\begin{equation}
\displaystyle{ \frac{d \ r}{d \ \epsilon} = r , \quad \frac{d \ y}{d \ \epsilon} = y, \quad \frac{d \ u_{J}}{d \ \epsilon} = \sigma^J  } \ .
\label{derivas}
\end{equation}
From this system of equations, we can define the flow of $\bf{v}$ on the infinite jet space to be the solution of the above  system with given initial values $(x,u^{(\infty)})=(x^0,u^0,u_J^0)$, $\ u^0 \equiv f(r,y)$ being  a solution of the  system of equations (\ref{systemN}):
\begin{equation}
exp\left[\epsilon \ pr \textbf{v}\right](x,u^{(\infty)})=\left(x(\epsilon),u^{(\infty)}(\epsilon)\right) \ .
\end{equation}

In our case, from the equation (\ref{derivas}) we must handle  the following system of coupled differential equations (we only write the actually relevant equations for our purpose, which is no  more than obtaining the flow on the manifold $M$):
\begin{eqnarray}
&&\displaystyle{\frac{d r}{d \epsilon} = r, \qquad  \frac{d y}{d \epsilon}= y} \label{flujo1}\\
&&\displaystyle{\frac{d u}{d \epsilon}= -u-\sum_{n=2}^N h_n(y) u_y^{(n}} \label{flujo2}\\
&&\displaystyle{\frac{d u_y^{(k}}{d \epsilon}= -(k+1) u_y^{(k}-D_y^{(k}\left[\sum_{n=2}^N h_n(y) u_y^{(n}\right]} \ ,
\label{flujojet}
\end{eqnarray}
where the notation $u_y^{(k}\equiv \partial_y^k u $ has been used.
The first two equations in (\ref{flujo1}) lead to the already known transformation of the independent variables $\displaystyle{\left(r = e^{\epsilon} r^0, y = e^{\epsilon} y^0 \right) }$, whereas the following  ones (\ref{flujo2}), (\ref{flujojet}) turn out to be decoupled, as we will see now, when we take into account some considerations about the variable $u$ and the functions $h_n(y)$. In order to see this, let us consider  the above equations (\ref{flujojet}) when the total derivative $D_y^{(k}$ has been developed:
\begin{equation}
\displaystyle{\frac{d u_y^{(k}}{d \epsilon}= -(k+1) u_y^{(k} -\sum_{n=2}^N \sum_{j=0}^k {{k} \choose {j}} h_n^{(j} (y)  u_y^{(n+k-j}}  \ ,
\label{flujojet3}
\end{equation}
Let us consider now the derivatives appearing in the right hand side of the equation (\ref{flujojet3}). On the one hand, the derivatives of the functions $h_n(y)$ (which have been denoted by $h_n^{(j} (y)$) vanish for $j> n-2$ since the degree with respect to $y$ of these functions is $n-2$. And on the other hand, the variable $u$, which is a function of the independent variables $r$ and $y$, is a solution of the system of equations (\ref{systemN}) and therefore, $\partial_y^k u=0$, $\forall k> N$. Thus, the order of the derivatives of $u$ appearing in (\ref{flujojet3}) fulfills the condition $n+k-j\leq N$.
With these considerations, the non-vanishing terms of the sum of index $j$ in (\ref{flujojet3}) correspond to the following range of the index $j$:  $n+k-N\leq j\leq n-2$; and hence, that equation can be written as follows:
\begin{equation}
\displaystyle{\frac{d u_y^{(k}}{d \epsilon}= -(k+1) u_y^{(k} -\sum_{n=2}^N \sum_{j=n+k-N}^{n-2} {{k} \choose {j}} h_n^{(j} (y)  u_y^{(n+k-j}} \ , \label{flujosola}
\end{equation}

Moreover, for $k=N$ and $k=N-1$ in (\ref{flujosola}) the lower value of the index $j$ is higher than the upper value and, therefore, the only term in the right hand side of the corresponding differential equations of (\ref{flujosola}) is the first one, whereas for higher values of $k$, it is straightforward to see that  the corresponding differential equations can be written as follows: 
\begin{eqnarray}
\displaystyle{\frac{d u_y^{(N}}{d \epsilon}}&=&-(N+1) u_y^{(N}  \nonumber\\
\displaystyle{\frac{d u_y^{(N-1}}{d \epsilon}}&=&-N u_y^{(N-1}  \nonumber\\
\displaystyle{\frac{d u_y^{(N-a}}{d \epsilon}}&=& -(N-a+1) u_y^{(N-a} -\sum_{n=2}^N \sum_{j=n-a}^{n-2} {{N-a} \choose {j}} h_n^{(j} (y)  u_y^{(n+N-a-j} \nonumber \\
 &=&-(N-a+1) u_y^{(N-a} -\sum_{n=2}^N \sum_{j=0}^{a-2} {{N-a} \choose {n-a+j}} h_n^{(n-a+j} (y)  u_y^{(N-j}
\ , \nonumber \\
 \label{desacos} 
\end{eqnarray}
where $a$ takes values from $2$ to $N-1$.

These decoupled equations can be integrated successively with the given initial values previously mentioned $(x,u^{(\infty)})=(x^0,u^0,u_J^0)$, and afterthat we can solve, with these solutions, the equation (\ref{flujo2}) which gives us the group transformation of the variable $u$.

\vspace*{2mm}
As a matter of illustration, let us consider for example the case $N=2$; i.e.,   let us try to solve the above equations (\ref{flujo2}), (\ref{desacos}) in order to obtain the action of the group of transformation $G^{(2)}_{M_2}$, which represents a symmetry of the system of equations (\ref{systemN}). The resulting equations for this case are as follows:
\begin{eqnarray}
&&\displaystyle{\frac{d u}{d \epsilon}=-u+\frac 13 u_{yy}}\label{casoN21} \\
&&\displaystyle{\frac{d u_y}{d \epsilon}=-2 u_y} \label{casoN22} \\
&&\displaystyle{\frac{d u_{yy}}{d \epsilon}=-3 u_{yy}} \ , \label{casoN23}
\end{eqnarray}
 and therefore the solution of (\ref{casoN23}), which is unique  derivative involved in (\ref{casoN21}), 
is given by:
\begin{equation}
\displaystyle{u_{yy}(x,\epsilon)=e^{-3 \epsilon} u_{yy}^{0}} \ ,
\end{equation}
with the initial condition $u_{yy}(x,0)=u^0_{yy}$. And the solution of equation (\ref{casoN21}) is
\begin{equation}
\displaystyle{u(x,\epsilon)=e^{-\epsilon} \left[ u^{0}+\frac{1}{6} u_{yy}^{0} (1-e^{-2 \epsilon})\right]} \ ,
\end{equation}
where $u(x,\epsilon=0)= u^0$ has been considered as the initial condition.

Thus, the one-parameter group generated by $\textbf{v}$ (\ref{uveN}) in $M$ for the case $N=2$ is given by:
\begin{equation}
\left(\tilde{x}, \tilde{u} \right)=\left[e^{\epsilon} r,\  e^{\epsilon} y,\  e^{-\epsilon} \left(u^0+\frac{1}{6} u_{yy}^0 (1-e^{-2 \epsilon})\right)\right] \ .
\label{action2}
\end{equation}

\section{Group-Invariant solutions}

A solution $u=f(x)$ of a system of differential equations is said to be  $G$-invariant if it remains unchanged by all  group transformations in $G$, meaning that for each element of the group  both the function $f$ and that transformed by the action of this element agree in their common domains of definition. If G is a symmetry  group of a system of differential equations $\bigtriangleup$, then we can find all  the $G$-invariant solutions to $\bigtriangleup$ by solving a reduced system  of equations, which will involve fewer  independent variables than the original system.

\subsection{$G_{MD}$-invariant solutions}

According to (\ref{action})  global invariants of the group $G_{MD}$ are $\kappa=y/r$ and $\mu=u y$, and hence a group-invariant solution takes the form:
\begin{equation}
u=\displaystyle{\frac{1}{y} f(\kappa)} \ .
\label{efe}
\end{equation}
Solving for the derivatives of $u$ with respect to $r$ and $y$ in terms of those of $f$ with respect to $\kappa$ and so on, and substituing those expressions in the Laplace equation, leads to the following linear, constant coefficient differential equation:
\begin{equation}
f^{\prime \prime} \kappa^2- 2 f^{\prime} \kappa + 2 f = 0 \ ,
\label{fk2}
\end{equation}
whose general solution is $f= a \kappa^2+b \kappa$. Therefore, the invariant solution of the Laplace equation by the action of  group $G_{MD}$ is simply the gravitational Monopole-Dipole solution:
\begin{equation}
u(x) = \displaystyle{a \frac{y}{r^2}+b \frac{1}{r}} \ .
\end{equation}

\vspace{2mm}

An alternative way to obtain  group-invariant solutions of the system of differential equations (\ref{systemN})
is as follows: First, we solve the supplementary equation, which is a PDE but which involves only one coordinate, its general solution being $u(r,y)=A(r) y+B(r)$.
Now, we make this solution  invariant under the action of $G_{MD}$ (\ref{action}); i.e., the transformed function $\tilde {u}(\tilde{r},\tilde{y})$ (\ref{trans})  functionally becomes the same with  $u(\tilde{r},\tilde{y})$, and therefore we may conclude that the only solutions for $A(r)$ and $B(r)$ are: 
\begin{equation}
\displaystyle{A(r)=\frac{a}{r^2} \qquad , \qquad B(r)=\frac{b}{r}} \ ,
\end{equation}
whose transformation by the action of the group proves to be:
\begin{equation}
\tilde{u}(\tilde{r},\tilde{y})=e^{-\epsilon}u(r,y)=e^{-\epsilon}\displaystyle{\left(\frac{a e^{-\epsilon} \tilde{y}}{\tilde{r}^2 e^{-2 \epsilon}}+\frac{b}{\tilde{r} e^{-\epsilon}} \right)= a \frac{\tilde{y}}{\tilde{r}^2}+b \frac{1}{\tilde{r}}} .
\end{equation}

As  seen, this method is equivalent to the previous one and proves to be easier to develop because we do not need either to search for the characteristics of the vector field or solve the resulting differential equation (\ref{fk2}) after substituing the global invariants and their derivatives in the Laplace equation.

\subsection{$G_{M_n}^{(N)}$-invariant solutions}

We  present now two equivalent methods to obtain the group-invariant solutions for the case of the generalized vector field $\textbf{v}$ (\ref{uveN}): firstly, by using  the  explicit expression for the group of transformations obtained by integrating (\ref{flujo2}),(\ref{desacos}) for any value of $N$; and secondly,  by using expressions (\ref{infini}-\ref{esto}) which outline the infinitesimal flow by means of a formal Lie series in terms of the characteristic of the evolutionary vector field. 

\vspace*{2mm }
\noindent {\bf A)} For the {\it first method} we proceed in analogy with the alternative way seen in the previous case of monopole-dipole symmetry. Let us  illustrate the method by considering, as an example,  the case $N=2$: we solve  the supplementary equation of  the system of equations (\ref{systemN}) i.e., $u\equiv f(r,y)=A(r)+B(r) y+C(r) y^2$  and require this solution to be invariant under the action of the group $G^{(2)}_{M_n}$ (\ref{action2}).

Since group $G^{(2)}_{M_n}$ was proved to be a symmetry group , if $u^0=f(x)$ is a solution of the system of differential equations (\ref{systemN}), so are the functions:
\begin{equation}
\displaystyle{\tilde{u}=e^{-\epsilon} u^0 = e^{-\epsilon} \left[ f(r,y) +\frac 16 f_{yy}(r,y) (1-e^{-2 \epsilon})\right]} \ .
\end{equation}
Thus, the transformed function $\tilde{u}(\tilde{r},\tilde{y})$ will be invariant under the action of the group if the following equations are fulfilled:
\begin{eqnarray}
A(\tilde{r})&=& e^{-\epsilon}\left[A(\tilde{r} e^{-\epsilon})+\frac 13(1-e^{-2\epsilon}) C(\tilde{r}e^{-\epsilon})\right]\\ 
B(\tilde{r})&=& e^{-2 \epsilon} B(\tilde{r} e^{-\epsilon})\\ 
C(\tilde{r})&=& e^{-3 \epsilon} C(\tilde{r} e^{-\epsilon}) \ ,
\end{eqnarray}
which have the following unique kind of solutions:
\begin{equation}
\displaystyle{A(r)=\frac{a}{r}-\frac 13 \frac{c}{r^3} \ , \quad B(r)=\frac{b}{r^2} \ . \quad C(r)=\frac{c}{r^3}}
\end{equation}
The resulting invariant solution turns out to be:
\begin{eqnarray}
u(r,y)\equiv f(r,y)&=& \displaystyle{\left(\frac{a}{r}-\frac 13 \frac{c}{r^3}\right)+ y \left(\frac{b}{r^2}\right)+\left(\frac{c}{r^3}\right)y^2}=  \nonumber\\
&&\displaystyle{\frac{a}{r}+\frac{b}{r^2}P_1(y)+\frac{2c}{3 r^3} P_2(y)} \ ,
\end{eqnarray}
the gravitational multipole solution possessing monopole, dipole and quadrupole moments.

\vspace*{2mm}
\noindent {\bf B)} The {\it second method} to obtain  group-invariant solutions has to be  with the action of the group defined in (\ref{actionvq}). When the formal Lie series (\ref{infini}), (\ref{esto}) was previously outlined, we  mentioned that  the complete action of the group is not nedeed to obtain group-invariant solutions, because an alternative procedure arises from those expressions. The argument is supported by the following:

\vspace*{2mm}
\noindent {\bf Proposition}:

The  functions $u=f(x)$, solutions of the system of differential equations $\bigtriangleup_{\nu}(x,u^{(n)})=0$ given by (\ref{systemN}), are group-invariant solutions by the action of $G_{M_n}^{(N)}$ (the symmetry  group  whose infinitesimal generator is given by the generalized vector field (\ref{uveN})) if and only if the characteristic of the associated evolutionary vector field  $\textbf{v}_Q$ is zero over those functions.

\vspace*{5mm}

\noindent Proof:

Since the component of the evolutionary vector field is simply the characteristic $Q[u]$, if a solution $u=f(x)$ is group-invariant this obviously means that $\textbf{v}_Q(u)=0$, and therefore $Q(x,u^{(n)})=0$. Conversely, from the previous expression (\ref{infini}-\ref{esto}) we see that the prolongation of the evolutionary vector and all its successive prolongations $(pr \textbf{v}_Q)^n$ depend on $Q$ and its total derivatives, and hence the action of the group  (\ref{infini}) on every solution $u=f(x)=u(x,0)$ that makes $Q\mid_u=0$ leads to the equality $u(x,\epsilon)=u(x,0)$, i.e. the solution is group-invariant because the group transforms the solution into itself.  $\hfill{\square}$

\vspace*{5mm}

By applying this proposition, we can obtain  group-invariant solution of the system of differential equations $\bigtriangleup_{\nu}(x,u^{(n)})=0$ (\ref{systemN}) as follows: We first solve $\bigtriangleup_2^{(N)}$ and demand that the characteristic of this solution  be zero. In fact, that condition, $Q\mid_u=0$, is exactly the corresponding homogeneous equation (\ref{rn}):
\begin{equation}
Q\mid_u= 0= -u- r\displaystyle{\partial_r u- y\partial_y u-\sum_{n=2}^N h_n(y)\partial^n_y u} \ .
\end{equation}
The general solution of supplementary equation $\bigtriangleup_2^{(N)}$ from (\ref{systemN}) is: 
\begin{equation}
u(r,y)=\sum_{k=0}^N F_k(r) y^k \ .
\label{uu}
\end{equation}
If we force that solution to make  the characteristic zero, we find that:
\begin{equation}
\sum_{k=0}^N \left(F_k y^k+F_k^{\prime} y^k r + k F_k y^k\right)+\sum_{n=2}^Nh_n(y) \partial_y^n u=0 \ .
\label{ecua}
\end{equation}
It is clear that
\begin{equation}
\partial_y^n u = \sum_{k=n}^N F_k \displaystyle{\frac{k!}{(k-n)!}} y^{k-n} \ ,
\end{equation}
and therefore the second sum in the expression (\ref{ecua}) turns out to be:
\begin{equation}
\sum_{n=2}^Nh_n(y) \partial_y^n u=\sum_{n=2}^N \left[\sum_{i=0}^{n-2} H_{ni} y^i\right] \sum_{p=0}^{N-n} F_{p+n}(r) \displaystyle{\frac{(p+n)!}{p!}} y^p \ ,
\end{equation}
where the following notation has been used for the functions $h_n(y)$ (\ref{haches}):
\begin{equation}
h_n(y) \equiv \sum_{i=0}^{n-2} H_{ni} y^i \ .
\end{equation}

If we rearrange the sums  in powers of the variable $y$ we have: 
\begin{equation}
\sum_{n=2}^Nh_n(y) \partial_y^n u=\sum_{k=0}^{N-2}y^k \left[\sum_{j=0}^{k} \sum_{p=k+2}^{N} F_{p}(r) H_{p-k+j \ j} \displaystyle{\frac{p!}{(k-j)!}}\right] \ .
\end{equation}
The equation for the characteristic (\ref{ecua}) is now:
\begin{equation}
\sum_{k=0}^N  y^k\left[ (k+1) F_k +F_k^{\prime} r\right] + \sum_{k=0}^{N-2}y^k \left[\sum_{j=0}^{k} \sum_{p=k+2}^{N} F_{p}(r) H_{p-k+j \ j} \displaystyle{\frac{p!}{(k-j)!}}\right] =0 ,
\end{equation}
and so for each power in the variable $y$ we have the following equations:
\begin{eqnarray}
(k+1) F_k +F_k^{\prime} r=0  \qquad  , \qquad k=N, N-1  &\nonumber\\
(k+1) F_k +F_k^{\prime} r= -\sum_{j=0}^{k} \sum_{p=k+2}^{N} F_{p}(r)  H_{p-k+j \ j} \displaystyle{\frac{p!}{(k-j)!}},& \nonumber\\ 
k=0, \cdots , N-2 &\ ,
\label{solus}
\end{eqnarray}
whose solutions provide us with the corresponding functions for the potential $u$ (\ref{uu})  as follows:
\begin{eqnarray}
F_N(r) &=& \displaystyle{\frac{c_N}{r^{N+1}}} \nonumber\\
F_{N-1}(r) &=& \displaystyle{\frac{c_{N-1}}{r^{N}}} \nonumber\\
F_k(r) &=& \displaystyle{\frac{c_k}{r^{k+1}}-\frac{1}{r^{k+1}}\sum_{j=0}^k\frac{1}{(k-j)!} \sum_{p=k+2}^N H_{p-k+j \ j} \int F_p(r) r^k \ dr}  \nonumber\\
& & k=0, \cdots , N-2 \ .
\label{efes}
\end{eqnarray}

Let us finish this section with an example. We look for the $G^{(3)}_{M_n}$ group-invariant solutions of the  Laplace equation. From the above expressions (\ref{efes}) we have that:
\begin{equation}
F_3(r)=\displaystyle{\frac{c_3}{r^4}} \qquad , \qquad F_2(r)=\displaystyle{\frac{c_2}{r^3}}
\end{equation}
\begin{equation}
F_1(r)=\displaystyle{\frac{c_1}{r^2} -\frac{3!}{r^2} \left[ (H_{20} +H_{31}) \int \frac{c_3}{r^3} \ dr \right] = \frac{c_1}{r^2}-\frac{3}{5} \frac{c_3}{r^4}} \ ,
\end{equation}
which is the general solution of the equation $\displaystyle{2 F_1(r)+F_1^{\prime}(r) r -\frac 65 \frac{c_3}{r^4} = 0}$, and for $k=0$ in (\ref{efes}), we have:
\begin{equation}
F_0(r)=\displaystyle{\frac{c_0}{r} -\frac{2!}{r} \left[ H_{20} \int  \frac{c_2}{r^3} \ dr \right] = \frac{c_0}{r}-\frac 13 \frac{c_2}{r^3}} \ ,
\end{equation}
which is  the general solution of the equation $\displaystyle{F_0(r)+F_0^{\prime}(r) r -\frac 23 \frac{c_2}{r^3} = 0}$.

Finally, the invariant solution $u(r,y)$ is:
\begin{eqnarray}
u(r,y) &=& \displaystyle{\left(\frac{c_0}{r}-\frac 13 \frac{c_2}{r^3}\right)+ y \left(\frac{c_1}{r^2}-\frac 35 \frac{c_3}{r^4}\right)+ y^2 \left(\frac{c_2}{r^3}\right)+y^3\left(\frac{c_3}{r^4}\right)}= \nonumber\\
&&\displaystyle{\frac{c_0}{r}+\frac{c_1}{r^2}+\frac 23 c_2\frac{P_2(y)}{r^3}+\frac 25 c_3\frac{P_3(y)}{r^4}} \ ,
\end{eqnarray}
i.e., the partial sum of order $4$ of series (\ref{serie}) representing the gravitational solution constructed with  the mass, the dipole, the quadrupole and the octupole multipole moments.

\section{Conclusion}

The gravitational multipole solutions of  Newtonian gravity are well known, and of course it is not the aim of this work to discover them but to introduce  new insights into the mathematical description of the solutions of the gravitational potentials with a prescribed number of multipole moments. For the case of Newtonian gravity, it is known that these solutions are given by the partial sums of the series (\ref{serie}). In this work, we have shown that these solutions can be viewed as group-invariant solutions of certain one-parameter groups of transformations $G_{MD}$ (monopole-dipole case) and $G^{(N)}_{M_n}$ (general case).

The result obtained is a kind of generalization of the fact that the Newtonian Monopole is defined by the solution with spherical symmetry of the Laplace equation. The existence of some symmetries of the Laplace equation is proved, which allows us to restrict the solutions of that equation to those with the prescribed MM. In order to do so, it is necessary to append a so-called supplementary equation to the Laplace equation in order  to provide a system  that will admit the defined group as a symmetry.

The infinitesimal generators of the group as well as the action of the group have been constructed for each symmetry (in the general case we calculate specifically the action of the group $G_{M_n}^{(2)}$ to illustrate the procedure). And finally it is proved that the group-invariant solutions are  exactly the gravitational multipole solutions.

The supplementary equation arises as a constraint imposed on the Laplace equation in order to maintain the symmetry of the system of both equations, but this restriction itself does not  limit the solution of the  Laplace equation  to those desired gravitational multipole solutions if in addition one does not demand  an asymptotically well-behaved condition at infinity. For example, for the case $N=2$, the group $G^{(2)}_{M_n}$ is a symmetry of  the system of equations (\ref{systemN}) whose general solution is:
\begin{equation}
u(r,y)= \displaystyle{a_2-\frac{a_1}{r}+\frac{b_2 y}{r^2}+\frac{2 c_2 P_2(y)}{3 r^3}+r b_1 y+\frac 23 r^2 c_1 P_2(y)} \ ,
\label{gscaso2}
\end{equation}
where the last two terms must be rejected for asymptotic reasons. 

Alternatively,  both methods proposed to  obtain the group-invariant solutions for the general case (subsections $5.2 A$ and $5.2 B$)  lead univocally to the specific gravitational multipole solutions. In particular, first method provides a smart and mathematically standard procedure, with the additional advantage that we are dealing with  an algebraic condition that is easier to solve than a differential one\footnote{Note that the general solution above expressed (\ref{gscaso2}) requires us to solve, for instance, the following coupled differential equations:
\begin{eqnarray}
&u(r,y)\equiv A(r)+B(r) y+C(r) y^2 \nonumber\\
&2r A^{\prime}(r)+r^2 A^{\prime\prime}(r)+2 C(r)=0 \nonumber\\
&2r B^{\prime}(r)+r^2 B^{\prime\prime}(r)-2 B(r)=0 \nonumber\\
&2r C^{\prime}(r)+r^2 C^{\prime\prime}(r)-6 C(r)=0 \nonumber
\end{eqnarray}}. Of course  we already know the solution of the axially symmetric Laplace equation and the interpretation of the truncated series from (\ref{serie}), but the theoretical  result  given by  the existence of some extra symmetries of the Laplace equation is that it provides us with a procedure to obtain the gravitational multipole solutions with prescribed multipole moments without solving that differential equation.

A  clearly more relevant feature of these results  has to be stressed in the sense that they serve as a trial for a future generalization to General Relativity (GR). Apart from the fact that the non-linearity of GR gravitation does introduce calculation problems, the static and axisymmetric-vacuum metrics are described by two metric functions; one of them (which provides the other one by means of a quadrature) is actually a solution of the  Laplace equation. Whereas  multipole moments in Newtonian gravity provide  physical meaning to the coefficients arising from the general solution series (\ref{vacuum}), the GR scenario requires complicated computations to obtain the relation between those coefficients and Relativistic Multipole Moments (RMM) \cite{mio2}, \cite{geroch}, \cite{hansen}.

Some authors have devoted much effort in  researching  techniques aimed at obtaining solutions of stationary axisymmetric vacuum solutions of  Einstein field equations with prescribed multipole moments. In \cite{mio}, \cite{mio2} the Monopole-Quadrupole solution was obtained and  \cite{mio3} addresses the  MJQ approximate solution. In \cite{mio2} the authors obtained algebraic conditions $a_n=a_n(M_n)$ relating the coefficients of the series (\ref{vacuum}) with the RMM, and a metric function that is a solution of the Laplace equation is given explicitly. In \cite{sueco1}, \cite{sueco2} the authors developed a very interesting and useful method for generating the coeffcients $a_n$  needed to construct the gravitational multipole solution with prescribed RMM. Since the existence and uniqueness of that kind of solution in GR can be proved \cite{sueco2}, it would be an exceptional goal to establish  the existence of some kind of "symmetry" that would generalize  Birkhoff's theorem for a solution with a given set of RMM. The uniqueness theorems of partial differential equations are in general  very difficult  to solve and a topic of research in Mathematical Physics, and they would not strictly be the aim of future works. However, the proposal   outlined here consists of considering the algebraic relations $a_n=a_n(M_n)$ from a different mathematical point of view, hopefully searching for the existence of some kind of  "symmetry" rather than a boundary condition problem from which those relations arose.

Since it is Laplace's equation that we wish to solve, the problem could be oriented towards  finding the appropriate system of coordinates or some basis of functions where the already known gravitational multipole solutions, as in the case of Newtonian gravity, could be considered as the only solutions satisfying suitable symmetry conditions.

\section{Appendix A}
Let us consider the newtonian gravitational potential of a mass distribution with density
$\rho(\vec z)$, given by the following solution of the Poisson equation
\begin{equation}
\Phi (\vec x) = -G \int_V\frac 1R \rho(\vec z) d^3\vec z \qquad ,
\label{tres}
\end{equation}
where $G$ is the gravitational constant; the integral is extended to the volume of the source; $\vec
z$ is the vector that gives the position of a generic point inside the source, and 
$R$ is the distance between that point and any exterior point
$P$ defined by its position vector $\vec x$. Let us now make an expansion of this potential in  a power series of the inverse of the distance from the origin to the point $\vec P$ ($r \equiv
{\cal j}\vec x {\cal j}$) by means of a Taylor expansion\footnote{see \cite{tesis} for details} of the term  $\displaystyle{\frac{1}{R}}$ around the origin of coordinates, where $R\equiv \sqrt {(x^i-z^i)(x_i-z_i)}$.   The result is the followingç:
\begin{equation}
\Phi(\vec x) = -\frac{GM}{r}-G \sum_{l=1}^{\infty} \frac{1}{l!}
\frac{1}{r^{l+1}} Q^{i_1 \dots i_l} n_{i_1} \dots n_{i_l} \qquad ;
\qquad n^i \equiv \frac{x^i}{r} \qquad ,
\label{cuatro}
\end{equation}
where $M$ represents the total mass of the source, i.e.,  
\begin{equation}
M \equiv \int_V \rho(\vec z) d^3 \vec z \qquad ,
\label{cinco}
\end{equation}
and the quantities $Q^{i_1 \dots i_l}$ are completely symmetric and trace-free tensor fields defined as follows:
\begin{equation}
Q^{i_1 \dots i_l} \equiv (2l-1)!! {\cal T} \int_V z^{i_1} \dots
z^{i_l} \rho(\vec z) d^3 \vec z \qquad ,
\label{seis}
\end{equation}
where the symbol ${\cal T}$ denotes the subtraction of the traces.

By using spherical harmonic functions it is possible to give a better representation of the multipole moments  from the above-mentioned expansion of the function $\displaystyle{\frac{1}{R}}$ if it is written in terms of the Legendre polynomials as follows:
\begin{equation}
\frac 1R = \sum_{l=0}^{\infty} \frac 1{r^{l+1}} z^l P_l(\cos\beta)
\qquad ,
\label{once}
\end{equation}
where  $z \equiv {\cal j} \vec z {\cal j}$ and $\beta$ is the angle defined between the position vector of the exterior point $\vec x$ and the position vector of the interior point of integration $\vec z$. 

Let us write the Legendre polynomials in terms of the spherical harmonics:
\begin{equation}
P_l(\cos\beta) = \frac{4\pi}{2l+1} \sum_{m=-l}^{m=l}
Y_l^m(\theta,\phi) Y^{*m}_l(\hat\theta,\hat\phi) \qquad ,
\label{doce}
\end{equation}
where $(\theta,\phi)$ and $(\hat\theta,\hat\phi)$ are the angular spherical coordinates of the position vectors $\vec x$ and $\vec z$
respectively. Thus, the gravitational potential $\Phi(\vec
x)$ generated by  a mass distribution outside the source is given by
\begin{equation}
\Phi(\vec x) = -\frac{GM}{r} - 4\pi G\sum_{l=1}^{\infty} \frac
1{2l+1} \frac 1{r^{l+1}} \sum_{m=-l}^{m=l} D_l^mY_l^m(\theta,\phi)
\qquad ,
\label{trece}
\end{equation}
$D_l^m$ being some quantities defined as follows:
\begin{equation}
D_l^m = \int_V z^lY^{*m}_l(\hat\theta,\hat\phi)\rho(\vec z) d^3\vec
z \quad .
\label{catorce}
\end{equation}
For each value of $l$, there exist $2l+1$ complex quantities
$D_l^m$. Nevertheless, since $Y_l^{-m} = (-1)^m Y^{*m}_l$, only 
$2l+1$ real quantities are independent.

From expressions (\ref{trece}) and (\ref{cuatro}) it is clear that the completely symmetric and trace-free tensors $Q^{i_1 \dots i_l}$ are equivalent to the $2l+1$ real independent quantities \footnote{A completely symmetric tensor $T_{i_1 \dots i_r}$ of rank $r$,  in a manifold of dimension $3$ only has 
$(r+2)(r+1)/2$ independent components. In addition, the trace-free condition, i.e., $T_{i_1 \dots i_r} g^{i_1 i_2} = 0$ imposes $r(r-1)/2$ constraints, and therefore only $2r+1$ components of a completely symmetric and trace-free tensor  in dimension $3$ are independent.} $D_l^m$,
i.e., 
\begin{equation}
\frac 1{l!} Q^{i_1 \dots i_l} n_{i_1} \dots n_{i_l} =
\frac{4\pi}{2l+1} \sum_{m=-l}^{m=l}D_l^mY_l^m(\theta,\phi) \quad .
\label{quince}
\end{equation}

Finally, if we consider axial symmetry the number of quantities needed to define the multipole moments decreases. The only spherical harmonic that provides a non-vanishing integral in the expressions of $D_l^m$ is the corresponding $Y_l^0$, since the solution does not depend on the azimuthal angle. Therefore, only one quantity  remains to define the multipole moment of order $l$:
\begin{equation}
M_l = 2 \pi \int\int z^{l+2}\rho(z,\hat\theta)P_l(\cos\hat\theta)
\sin\hat\theta d\hat\theta dz \qquad ,
\label{dieciseis}
\end{equation}
$\rho(z,\hat\theta)$ being the density of the axially symmetric distribution; the integral is extended to the volume of the source, and therefore  $z$ represents the radius of the integration point and  
$\hat\theta$ the corresponding polar angle.

\section{Appendix B}
The functions $h_n(y)$ of equation (\ref{rn}) have been defined by (\ref{hach}) in terms of some quantities $C_{k,n}(y)$ from (\ref{erek}). By using formula (\ref{deriy}) in the recurrence relation (\ref{erres}) we immediately get
\begin{equation}
C_{k+1,n}(y)=C_{k,n}(y)-\frac{C_{k,k}(y)}{(2k-1)!!} \partial_y^k P_n(y) \quad , \quad n > k \quad . 
\label{ecces}
\end{equation}
This recurrence relation is solved by just plugging it into itself as many times as needed, but nevertheless it is not necessary to solve all the quantities of these double-index functions $C_{k,n}$, since we are only looking for the quantities $C_{n,n}$. As has been said, by developing the recurrence relation (\ref{ecces}) (with increasing value of $k$) starting from $C_{2,n}$ we easily see that
\begin{equation}
C_{k,n}(y)=C_{2,n}(y)-\sum_{j=2}^{k-1} \frac{C_{j,j}(y)}{(2j-1)!!} \partial_y^j P_n(y) \quad , \quad n \geq k \quad . 
\label{eccesbis}
\end{equation}
Finally, by taking $k=n$ and considering (\ref{hach}) and $C_{2,n}=-\partial_y P_{n-1}(y)$ we get the following relation to calculate the functions $h_n(y)$:
\begin{equation}
h_n(y)=\frac{-1}{(2n-1)!!} \left[ \partial_y P_{n-1}+\sum_{k=2}^{n-1}h_k(y)\partial_y^k P_n(y)\right]  \ , \quad n\geq 3 , \qquad h_2(y)=-\frac 13.
\label{haches}
\end{equation}

From these relations (\ref{haches}) we can  see that the functions $h_n(y)$ are polynomials in the variable $y$ of degree $n-2$:
\begin{eqnarray}
h_2(y)&=&-\frac{1}{3!!} \nonumber\\
h_3(y)&=&\frac{2}{5!!} y \nonumber\\
h_4(y)&=&-\frac{1}{7!!} (1+4y^2) \nonumber\\
h_5(y)&=&\frac{2}{9!!}(3y+4y^3) \nonumber\\
h_6(y)&=&-\frac{2}{11!!}(1+12y^2+8y^4)
\label{hachexp}
\end{eqnarray}

\section*{Acknowledgments}
This work was partially supported by the Spanish Ministry of Education and Science
 under Research Project No. FIS 2006-05319.
This work would not have been possible  without the ideas provided by Professor Jes\'us Mart\'\i n, whom I  wish to thank for helpful discussions spread out over a good number of years. 
I also wish to thank   Dr. Javier Villaroel for his 
comments about the bibliography and Dr. Alberto Alonso and Dr. Miguel \'Angel G\'onzalez-Leon for very fruitful insights.

\end{document}